\documentclass[
 reprint,
 bibnotes,
 amsmath,
 amssymb,
 aps,
 prb,
]{revtex4-2}

\usepackage{graphicx}
\usepackage{dcolumn}
\usepackage{bm}
\usepackage{braket}

\usepackage{hyperref}

\usepackage[mathlines]{lineno}

\usepackage{xcolor}

\begin{document}

\preprint{arXiv:}

\title{Infrared Magneto-Polaritons in MoTe$_2$ Mono- and Bilayers}

\author
{Bo Han,$^{1}$ Jamie M. Fitzgerald,$^{2}$ Lukas Lackner,$^{1}$ Roberto Rosati,$^{2}$ Martin Esmann,$^{1}$ Falk Eilenberger,$^{3,4,5}$ Takashi Taniguchi,$^{6}$ Kenji Watanabe,$^{7}$ Marcin Syperek,$^{8}$ Ermin Malic,$^{2}$ and Christian Schneider$^{1,}$}

\email
{Corresponding author: \\christian.schneider@uni-oldenburg.de}

\affiliation
{
$^{1}$Institut für Physik, Fakultät V, Carl von Ossietzky Universität Oldenburg, 26129 Oldenburg, Germany\\ 
$^{2}$Department of Physics, Philipps-Universität Marburg, Renthof 7, D-35032 Marburg, Germany\\
$^{3}$Fraunhofer-Institute for Applied Optics and Precision Engineering, 07745 Jena, Germany\\
$^{4}$Institute of Applied Physics, Abbe Center of Photonics, Friedrich Schiller Universität Jena, 07745 Jena, Germany\\
$^{5}$Max Planck School of Photonics, 07745 Jena, Germany\\
$^{6}$International Center for Materials Nanoarchitectonics, National Institute for Materials Science,
Tsukuba 305-0044, Japan\\
$^{7}$Research Center for Functional Materials, National Institute for Materials Science, Tsukuba 305-0044, Japan\\
$^{8}$Department of Experimental Physics, Faculty of Fundamental Problems of Technology, Wrocław University of Science and Technology, Wyb.Wyspiańskiego
27, 50-370 Wrocław, Poland
}
\date{\today}

\begin{abstract}
MoTe$_2$ monolayers and bilayers are unique within the family of van-der-Waals materials since they pave the way towards atomically thin infrared light-matter quantum interfaces, potentially reaching the important telecommunication windows. Here, we report emergent exciton-polaritons based on MoTe$_2$ monolayer and bilayer in a low-temperature open micro-cavity in a joint experiment-theory study. Our experiments clearly evidence both the enhanced oscillator strength and enhanced luminescence of MoTe$_2$ bilayers, signified by a 38 \% increase of the Rabi-splitting and a strongly enhanced relaxation of polaritons to low-energy states. The latter is distinct from polaritons in MoTe$_2$ monolayers, which feature a bottleneck-like relaxation inhibition. Both the polaritonic spin-valley locking in monolayers and the spin-layer locking in bilayers are revealed via the Zeeman effect, which we map and control via the light-matter composition of our polaritonic resonances.  
\end{abstract}

\maketitle

\section{Introduction}
Two-dimensional semiconductors of transition metal dichalcogenides (TMDCs) are an excellent research platform for solid-state cavity quantum electrodynamics due to their strong light-matter interactions and intriguing spin-valley locking \cite{wang2018colloquium,schneider2018two,mueller2018exciton}. Their optical properties hinge on very robust valley excitons with binding energies of hundreds of meV \cite{mak2010atomically,chernikov2014exciton,perea2022exciton}. Composite quasi-particles such as exciton-polaritons in TMDC-microcavity systems can inherit physical properties from both the cavity modes and the optically active material, including the ultralight effective mass \cite{carusotto2013quantum}, magnetic responses \cite{larionov2010polarized,fischer2014anomalies}, and non-linearities due to exchange-correlation \cite{lerario2017room,amo2009superfluidity}, dipolar interaction \cite{louca2023interspecies}, phase space filling \cite{song2024microscopic} and moiré confinement \cite{zhang2021van}. Therefore, they are  remarkable systems to explore collective phenomena such as bosonic condensation \cite{anton2021bosonic,kasprzak2006bose}, coherent light emission\cite{shan2021spatial,paik2019interlayer,wurdack2022enhancing,zhao2021ultralow}, polariton blockade \cite{delteil2019towards,munoz2019emergence} and correlated magnetism \cite{scherzer2024correlated}.  

Among the TMDC family members, MoTe$_2$ features a unique band structure with both the monolayer (ML) and bilayer (BL) exhibiting a direct band gap \cite{ruppert2014optical,aslan2018probing,helmrich2018exciton,robert2016excitonic,lezama2015indirect,kutrowska2022exploring}. The ground-state excitonic resonances lie in the near infrared spectral range $\sim$1.1 $\mu$m \cite{han2018exciton,kutrowska2022exploring}, making MoTe$_2$ a good candidate for optoelectronic \cite{zhao2021site} and quantum optical applications \cite{kavokin2022polariton} in the optical telecommunication window \cite{ramzan2023strained}. The extraordinary electronic properties \cite{jindal2023coupled,pan2023spin} also render MoTe$_2$ a promising material for studies of integer and fractional (anomalous) (spin) quantum Hall effects \cite{xu2023observation,cai2023signatures,park2023observation,kang2024evidence}. However, the strong light-matter coupling regime of MoTe$_2$ has not received notable attention so far compared to other TMDC materials.. 

In this work, we demonstrate the first experimental measurements on the strong coupling of excitons in MoTe$_2$ ML and BL with discretized cavity modes in a low-temperature open optical micro-cavity. Our study reveals that the BL features an increase in Rabi splitting and coupling strength by 38 \% compared to the ML, while maintaining otherwise identical conditions. Interestingly, we observe a bottleneck-like inhibition of relaxation in ML structures, effectively enhancing the luminescence from the upper polariton branch (UPB), while, in stark contrast, the majority of population relaxes to the lower polariton branch (LPB) in the BL. Our experiments are complemented by a many-particle theory revealing the microscopic mechanisms behind this qualitatively different behaviour in MoTe$_2$ ML and BL. Using systematic magneto-optics measurements, we explore the polaritonic Zeeman effect and reveal g-factors which are controlled by the cavity-exciton detuning, manifesting the valley and layer degree of freedom in the ML and BL, respectively. We also discover an unconventional enhancement of the coupling strength of both the ML and BL in the magnetic field.

\section{Strong Coupling and quasi-particle relaxation}
The MoTe$_2$ ML and BL are encapsulated \cite{castellanos2014deterministic} with thin hexagonal boron nitride flakes and deposited on a distributed Bragg reflector (DBR) with a stop band centered at 1050 nm (see Fig. S1(a,b,e) of Supplementary Material \cite{supp}). The samples are loaded in a closed-cycle low-temperature cryostat (3.5 K) equipped with a superconducting magnet. First, we perform the photoluminescence (PL) and differential reflectivity (DR) measurements on the encapsulated MoTe$_2$ flakes by using a linearly polarized 765 nm continuous wave laser and a tungsten-halogen lamp, respectively. The indiscernible Stokes shift between the PL and DR proves the high quality of our samples (see Fig. S1(f-i) \cite{supp}). More experimental details are summarized in Section III of \cite{supp}. 

\begin{figure}[t]
\includegraphics[width=\columnwidth]{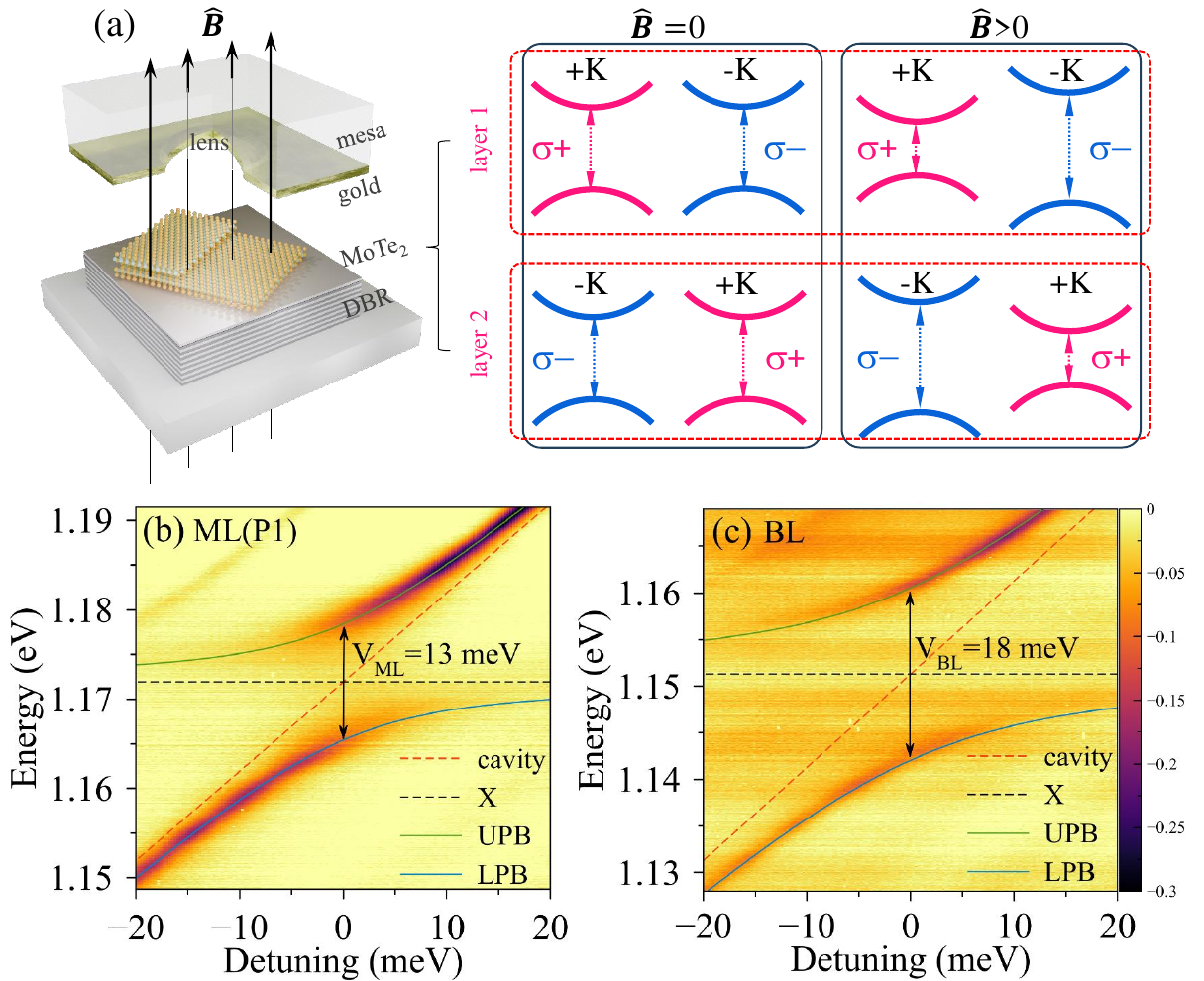}
\caption{\label{figure_1} (a) Schematic of the open cavity embedded with MoTe$_2$ ML and BL in magnetic field. The valley Zeeman effect for layers with inversion symmetry is sketched on the right panel. (b) DR as a function of cavity detuning with a cavity loaded with a ML (P1), featuring exciton energy $E_X^{ML}$=1.1719 eV and a Rabi-splitting of $V_{ML}^{0T}$=13 meV. (c) DR in BL as a function of cavity detuning, featuring $E_X^{BL}$=1.1513 eV and $V_{BL}^{0T}$=18 meV. The colorbar is identical for (b) and (c).}
\end{figure}

For strong coupling measurements, we use a top mirror that is a gold-coated silica mesa with a pre-manufactured hemispheric lens structure of 6 $\mu$m diameter and 300 nm depth (see Fig. S1(c,d) \cite{supp}). A schematic of the cavity setting is shown in Fig. \ref{figure_1}(a).  The lens and bottom DBR form discretized Tamm-plasmon lens modes (Q-factors $\sim$400) that can be readily measured in PL or DR. The focused excitation spots are comparable to the lens diameter. Our previous works have demonstrated the capability of discretized Tamm modes in tuning the emitters' spontaneous emission rates in the weak coupling regime \cite{drawer2023monolayer,han2023controlling}. DC voltage applied on the actuator can change the cavity length with sub-nm positioning resolution. The PL and DR measurements are thus performed with a fine detuning step size of 0.1 meV. The detuning $\Delta=E_C-E_X$ is defined as the energetic difference between the cavity and exciton modes. Both measured ML and BL samples are deliberately tuned to couple with the 8th longitudinal lens modes, corresponding to a cavity length of approximately 4.28 $\mu$m at $\Delta=0$. 

\begin{figure*}[t]
\includegraphics[width=1.8\columnwidth]{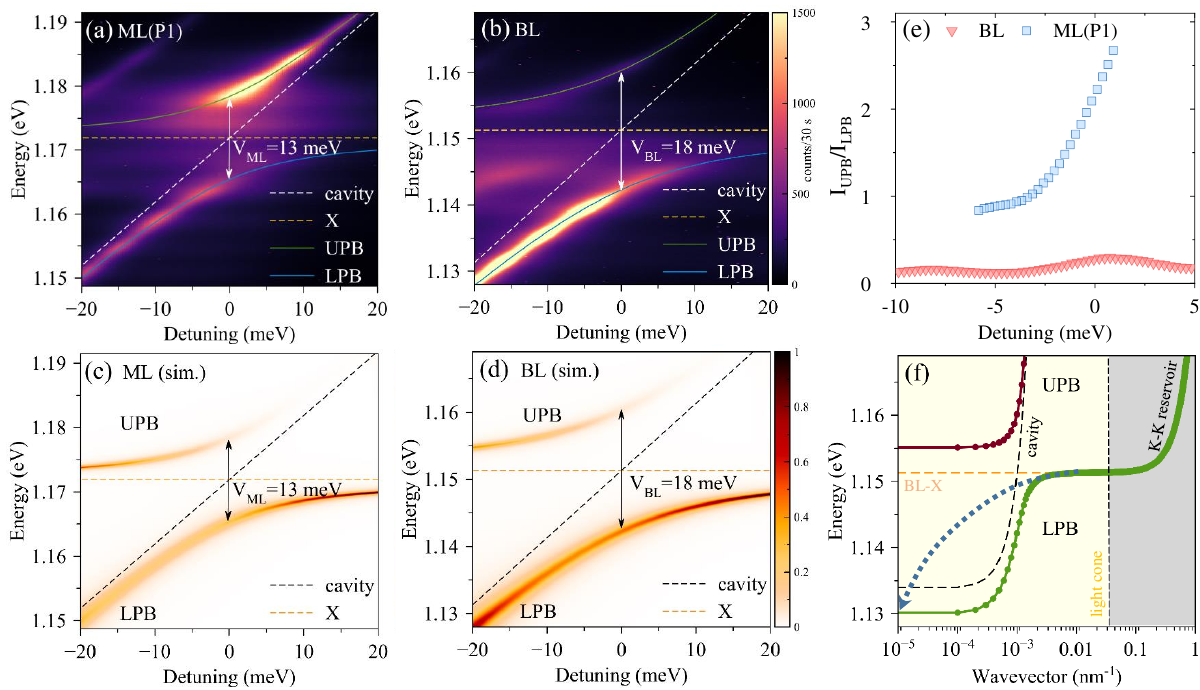}
\caption{\label{figure_2} PL spectra of exciton-polaritons in (a) ML (P1) and (b) BL as a function of detuning. The ML polaritons feature a significant PL emission from the UPB, whereas BL polaritons exhibit significant particle relaxation into the LPB. The coupled oscillator models for DR results as in Fig. \ref{figure_1} are also superimposed on the PL anti-crossings with excellent consistency. Theoretical description of the PL from (c) MoTe$_2$ ML- and (d) BL-polaritons, based on the material-realistic parameters from the experiments. Both absolute intensities are normalized to the BL emission, sharing the same colorbar of (d). (e) Experimentally extracted PL intensity ratio between the UPB and LPB in MoTe$_2$ ML and BL as a function of the detuning. (f) Schematics of the polariton relaxation regime. The detuning is set to -17.5 meV to have the LPB (K=0) lying 21 meV below the BL exciton, matching the energy of the homopolar A$_1$ optical phonons. The coupling strength and exciton energy at K=0 are experimental values. The opening of relaxation to LPB (K=0) is marked by the blue dashed arrow.}
\end{figure*}

In cavity-length dependent DR measurements in Fig. \ref{figure_1}(b,c), we observe clear anti-crossing features. We apply coupled oscillators theory to calculate the energy of both polariton branches as a function of the cavity-exciton detuning via  $E_{UPB(LPB)}$=$(E_X+E_C\pm\sqrt{V^2+\Delta^2})/2$. The fitting of polariton resonances in ML and BL are superimposed on the experimental DR results in Fig. 1(b,c), as well as in the PL measurements in Fig. \ref{figure_2}(a,b). This approach yields a coupling strength of $V_{ML}^{0T}$=13 meV for the MoTe$_2$ ML (position 1) and $V_{BL}^{0T}$=18 meV for the BL. The superscript 0 T indicates the absence of an externally magnetic field. We encounter a slight variation of the coupling strength for different sample positions of the ML on the order of the polariton linewidth (see Fig. S5-S6 for measurements on different positions: P2 and P3 \cite{supp}). Indeed, the observed increase of the Rabi-splitting by a factor of 1.38 for the BL is in agreement with the Tavis-Cummings model predicting a scaling with the square root of the number of oscillators \cite{rupprecht2020demonstration,krol2019exciton,tavis1968exact}.

A striking difference between ML and BL polaritons becomes obvious when comparing the PL intensity between UPB and LPB for the two cases. Figure \ref{figure_2}(a) depicts the PL from the polaritonic modes for the ML sample (P1), whereas the BL case is plotted in Fig. \ref{figure_2}(b) with identical excitation conditions. For BL-polaritons, the UPB emission is much weaker than the emission from the LPB, which is indicative of efficient relaxation of population to the energetically lower state. In stark contrast, for ML-polaritons, the UPB is generally more intense than the LPB (see Fig. S5(a) and Fig. S6(a) for other sample positions \cite{supp}). We plot the emission intensity ratio between the UPB and LPB in Fig. \ref{figure_2}(e) to quantify the relaxation strength for all detunings: in the BL case, the intensity ratio is generally smaller than 0.4, which is indicative for enhanced PL from the LPB, as a consequence of efficient relaxation. However, in the case of the ML, the ratio exceeds unity even for very red detuning, and increases dramatically towards the zero and blue detuning regimes.

\begin{figure*}[t]
\includegraphics[width=1.875\columnwidth]{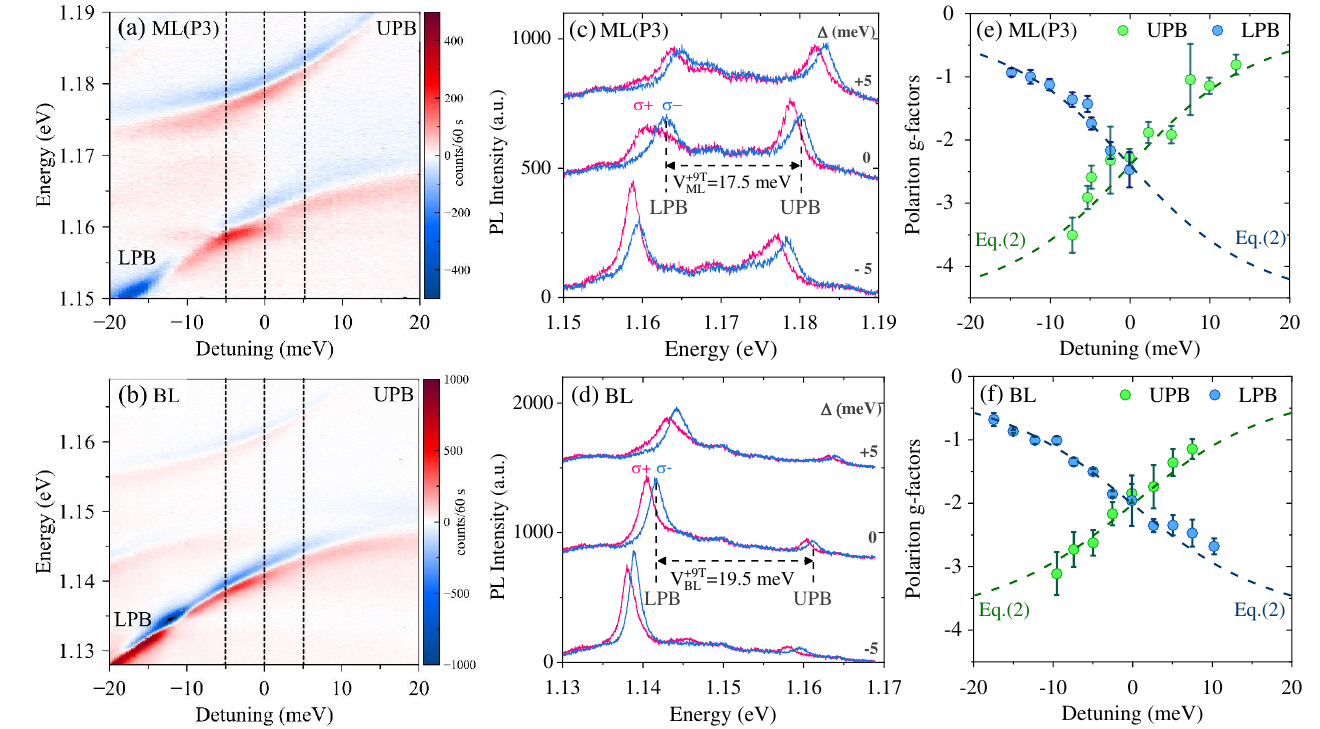}
\caption{\label{figure_3} Polariton Zeeman effect in spectral and polarization-resolved PL measurements at +9 T. The color-coded Zeeman plots in (a) ML (P3) and (b) BL are defined by subtracting the $\sigma-$ polarized spectrum from the $\sigma+$ polarized one at the same cavity detuning. (c) ML (P3) and (d) BL representative PL spectra of polariton Zeeman splits at $\Delta=$0 and $\pm$5 meV, corresponding to the dashed line marks in (a) and (b). The Rabi splittings are $V_{ML}^{+9T}$= 17.5 meV for the ML, and $V_{BL}^{+9T}$=19.5 meV for the BL. Experimental (dots) and theoretical (dashed lines) polariton g-factors in (e) ML (P3) and (f) BL. The error bars are derived by summing up the energetic error bars of the fitted polariton peaks in Fig. S5(c,e) \cite{supp}.}
\end{figure*}

We model the polariton population and relaxation in our systems using a material-realistic many-particle theory \cite{supp}. Exciton-polaritons are modelled within a Wannier-Hopfield framework \cite{fitzgerald2022twist}, using the experimentally extracted light-matter coupling parameters as well as a bare cavity linewidth of $1.5$ meV. Polariton-phonon scattering is described via the deformation potential and second-order Born-Markov approximation \cite{fitzgerald2024circumventing,brem2020phonon}. In particular, we explicitly account for the A$_1$ homopolar phonon with an energy of $21$ meV, which strongly interacts with excitons in MoTe$_2$ \cite{han2018exciton,sayers2023strong}. The values for the deformation potentials are extracted from experimentally measured temperature-dependent linewidths of the 1s exciton in both ML and BL MoTe$_2$ (see Fig. S7 \cite{supp}). We find an increased exciton-phonon coupling for the BL, in agreement with previous studies \cite{helmrich2018exciton}. The frequency-dependent PL at steady-state conditions is given by the polariton Elliot Formula \cite{fitzgerald2024circumventing}
\begin{equation}
I(\omega) = \sum_{\textit{n=\text{LPB,UPB}}} \frac{\gamma_n \Gamma_n}{(E_n-\hbar\omega)^2+(\gamma_n + \Gamma_n)^2} N_n^0
\label{Pol_intens}
\end{equation}
where $\gamma$ and $\Gamma$ are the polariton radiative and phonon-induced decay rates, respectively, $E_n$ is the polariton energy, and $N_n^0$ is the Boltzmann distribution of the quasi-particles. We fix the effective temperature of the exciton gas to $60$ K to consider the realistic case of imperfect thermalization. The results of the model are plotted for the case of ML- (Fig. \ref{figure_2}(c)) and BL-polaritons (Fig. \ref{figure_2}d)). Strikingly, the dramatic increase of the lower polariton PL for the BL is well captured in our microscopic description, which verifies the impact of the A$_1$ phonon mode on the relaxation dynamics. The more efficient polariton relaxation observed for the BL is attributed to the combined effect of the enhanced exciton-phonon coupling and larger Rabi splitting. The BL polariton has a larger excitonic character at the negative detunings when the scattering channel from the exciton reservoir opens up (see Fig. \ref{figure_2}(f)), i.e., $E_{\text{LPB}}+E_{\text{A$_1$}} \approx E_{\text{X}}$, where E$_{A_1}$ is the A$_1$ phonon energy. This, together with the larger exciton-phonon scattering in the BL, leads to a significantly larger occupation of the BL-LPB at all detunings.

\section{Polaritonic Magneto-optics}

Polaritons are light-matter hybrid quasi-particles whose magnetic responses stem from their excitonic nature. The polaritonic g-factor thus needs to be rescaled by the excitonic Hopfield coefficient $|X_k|^2$=$(\sqrt{\Delta_k^2+V^2}+\Delta_k)/2\sqrt{\Delta_k^2+V^2}$, a parameter characterizing the excitonic proportion in the polariton. However, the polaritonic Zeeman phenomena in TMDCs have not been systematically studied with different light-matter compositions.  

Due to time-reversal symmetry, the spin, orbital and valley indices in the ML take opposite signs between adjacent K valleys, leading to a locking of spin and valley degrees of freedom in bands with strong spin-orbit coupling. In contrast, the BL retains the inversion symmetry, which leads to the intriguing phenomenon of spin-layer locking \cite{jones2014spin,raiber2022ultrafast,brotons2020spin}. Both phenomena reveal distinctive spectral signatures in polarization spectroscopy in the presence of external magnetic fields, as sketched in Fig. 1(a). In order to extract the exciton out-of-plane g-factor, magneto-DR is firstly performed between $\pm$9 T in a Faraday geometry in the absence of the top mirror, ensuring weak coupling conditions. We utilize linearly polarized excitation, and measure counter-circularly polarized signals ($\sigma\pm$). The results are shown in Fig. S4 \cite{supp}. The out-of-plane excitonic g-factors ($g_{x}$) are calculated by fitting the Zeeman splitting of the valley excitons by using $\Delta$E=$g_{x}\mu_BB$, where $\mu_B$ is the Bohr magneton and $B$ is the magnetic flux density. The values of $g^{ML}_{X}$=-4.6 and $g^{BL}_{X}$=-4.1 are in excellent agreement with previous reports \cite{jiang2017zeeman,arora2017interlayer,goryca2019revealing}. 

We notice that the  modified sample position (P3), which is chosen for the magneto-optical study of the ML-polaritons features a slightly larger coupling strength of 16 meV measured at zero-field (see Fig. S3(a) and Fig. S5(a,b) \cite{supp}), which we attribute to locally varying dielectric screening and charging effects in the ML. Polaritonic Zeeman effects in both the MoTe$_2$ ML and BL are then extracted from magneto-PL measurements (Fig. \ref{figure_3}). We utilize the same polarization configuration for the polaritonic magneto-optics measurements. The polaritonic Zeeman splitting as a function of the detuning is then straightforwardly calculated via the energetic differences between $\sigma+$ and $\sigma-$ resonances at +9 T. 

We first plot the spectrally resolved detuning-dependent polarization patterns for ML and BL polaritons in Fig. \ref{figure_3}(a,b), which are obtained by subtracting the  $\sigma-$ polarized spectrum from the $\sigma+$ polarized one at the same cavity detuning. The same effects are also revealed in the color-coded magneto-DR difference spectra that are compiled in Fig. S5(d,f) \cite{supp}. We can immediately see the detuning dependent dichroism, which becomes especially prominent in the excitonic regime of the polariton branches, for example, at $\Delta=\pm20$ meV for the LPB and UPB, respectively. Magneto-PL and -DR difference spectra on another sample position (P2) of the ML, which display the same polaritonic Zeeman effects, are presented in Fig. S6 \cite{supp}. 

The PL spectra at detunings of 0 and $\pm$5 meV are also shown in Fig. \ref{figure_3}(c) for the ML case, and Fig. \ref{figure_3}(d) for the BL case.  The Zeeman splitting becomes very apparent in this representation, and acquires a similar magnitude for the ML and BL polaritons for comparable detunings. From the detuning dependent Zeeman effect, we can model the UPB and LPB g-factors as a function of detuning as:
 \begin{equation}
  g_{UPB(LPB)}=\frac{g_x}{2}\pm\frac{\sqrt{V^2+\Delta_1^2}}{2\mu_BB}\mp\frac{\sqrt{V^2+\Delta_2^2}}{2\mu_BB}
 \label{eq:g_UPB}
 \end{equation}
where $\Delta_1=E_C-E_X^{\sigma^+}$ and $\Delta_2=E_C-E_X^{\sigma^-}$ are the actual cavity detunings with respect to each excitonic Zeeman split, and $\Delta=(\Delta_1+\Delta_2)/2$. The zero-detunings of $\Delta_1$ and $\Delta_2$ are marked by double-sided arrows in Fig. S5(c,e) \cite{supp} that present the Lorentzian fitting results of the polariton Zeeman splits in ML and BL at +9 T. The experimentally extracted polaritonic g-factors of each branch in the ML (Fig. \ref{figure_3}(e)) and BL (Fig. \ref{figure_3}(f)) are in excellent agreement with our theoretically derived values from Eq. (\ref{eq:g_UPB}). The degeneracy of the valley as well as the layer-locked polaritons in the ML and BL is thus lifted, while the strong coupling effect is verified as a viable tool to tune the resulting optical dichroism.

It is furthermore worth noting that, comparing the coupling strength at +9 T (Fig. \ref{figure_3}(c,d)) to the scenario without the magnetic field (Fig. S3(b,h) \cite{supp}), the Rabi splittings are enhanced by 1.5 meV for both the ML and BL, corresponding to an effective enhancement of the coupling strength by approximately 8-9 \%. This behavior occurs consistently for various sample positions (see Fig. S6(e,f) \cite{supp}). The coupling strength enhancement by magnetic field was previously reported for polaritons in III-V semiconductor quantum wells embedded in monolithic cavities \cite{berger1996magnetic}, where the magnetic compression of exciton wavefunction increases the exciton oscillator strength and consequently enhances the vacuum-field Rabi splitting \cite{kavokin2017microcavities}. Our encounter is thus very unconventional, since the exciton Bohr radius ($a_B\sim$1-2 nm) is significantly smaller than the estimated magnetic length $L_B$=$\sqrt{\hbar/(e{\cdotp}B)}\approx$ 8.6 nm. 


\section{Conclusions}
We report the first optical microscopy measurements on the emergence of exciton-polaritons in MoTe$_2$ monolayer and bilayer. In contrast to the bilayer that exhibits an efficient population relaxation to the lower polariton branch, the relaxation in the monolayer features a pronounced bottleneck phenomenon, which has been modeled using a microscopic many-particle theory describing the scattering of exciton-polaritons with A$_1$ optical phonons. We have also verified the strong couplings via magneto-optics measurements, where the polariton valley and layer degeneracies are lifted in monolayer and bilayer, respectively. We can thus extract the  polaritonic g-factors as a function of the cavity-exciton detuning. Our work paves the way for further research involving cavity-mediated phenomena in MoTe$_2$-based van-der-Waals heterostructures, including the study of correlated phenomena, Telluride-based dipolaritons, and polariton lasers operated at telecommunication wavelengths. 

\begin{acknowledgments}
The project is funded by Deutsche Forschungsgemeinschaft (DFG) Lantern project (funding numbers: Schn1376 11.1). C.S. acknowledges DFG within the initiative for major equipment (Project INST184-220). B.H. acknowledges Alexander von Humboldt-Stiftung for the fellowship grant. M.E. acknowledges funding from the Carl von Ossietzky Universität Oldenburg through a Carl von Ossietzky Young Researchers' Fellowship. F.E. acknowledges support by DFG SFB 1375 (NOA) and BMBF FKZs 16K1SQ087K and 13XP5053A. J.F, R.R and E.M acknowledge funding from DFG via SFB 1083 and the regular project 524612380. K.W. and T.T. acknowledge support from the JSPS KAKENHI (grant nos. 21H05233 and 23H02052) and World Premier International Research Center Initiative (WPI), MEXT, Japan. M.S. acknowledges funding from the project No. 2019/35/B/ST5/04308 financed by the Polish National Science Center (NCN).  
\end{acknowledgments}

\bibliographystyle{apsrev4-2}
\bibliography{apssamp}

\end{document}